\documentclass[%
 reprint,
 amsmath,amssymb,
prb,
]{revtex4-1}
\usepackage{color}
\usepackage{graphicx}% Include figure files
\usepackage{dcolumn}% Align table columns on decimal point
\usepackage{bm}% bold math
\usepackage{hyperref}% add hypertext capabilities
\usepackage{natbib}
\usepackage{epstopdf}
\begin{document}

\preprint{APS/123-QED}

\title{Single-photon blockade in optomechanical photonic crystal cavity with third-order nonlinearity}

\author{Bijita Sarma}
 %\altaffiliation[Also at ]{Department of Physics, Indian Institute of Technology  Guwahati, Guwahati-781039, Assam, India}
 \email{s.bijita@iitg.ernet.in}
\author{Amarendra K. Sarma}%
 \email{aksarma@iitg.ernet.in}
\affiliation{%
 Department of Physics, Indian Institute of Technology Guwahati, Guwahati-781039, Assam, India}

%\collaboration{MUSO Collaboration}

%\affiliation{
%\affiliation{
%\affiliation{%

\date{\today}% It is always \today, today,

\begin{abstract}
Photon statistics in a weakly driven optomechanical photonic crystal cavity, with Kerr-type nonlinearity, is analyzed both analytically and numerically. Single-photon blockade effect is demonstrated via calculations of the zero-time-delay second-order correlation function $g^{(2)}(0)$. The analytical results obtained by solving the Schr\"{o}dinger equation are in complete conformity with the results obtained through numerical solution of the quantum master equation. The experimental feasibility with state-of-the-art device parameters is discussed. It is observed that photon blockade could be generated at the telecommunication wavelength, even at lower values of the nonlinear susceptibility parameters and in the weak single-photon optomechanical coupling regime.The system is found to be robust against pure dephasing-induced decoherences.

\begin{description}
\item[PACS numbers] 42.50.Ar, 42.65.-k,42.50.Lc,07.10.Cm
\verb||
\end{description}
\end{abstract}

\pacs{Valid PACS appear here}% PACS, the Physics and Astronomy
\maketitle

%\tableofcontents

\section{\label{sec:level1}INTRODUCTION}

On demand generation of single photon plays a central role in light-based quantum-information systems. Single-photon states \cite{englund2012ultrafast, kiraz2004quantum} are highly critical for applications in quantum computation, cryptography and metrology.\cite{knill2001scheme,duan2001long,kimble2008quantum,o2009photonic} Significant progress has been made in recent years towards demonstration of single-photon sources \cite{lounis2005single} in various platforms, such as single quantum dots integrated with photonic crystal cavities, \cite{toishi2009high, majumdar2012probing, faraon2011integrated, hennessy2007quantum, ishida2013photoluminescence} optical fibers \cite{xu2007plug, xu2008plug} or single atoms coupled with micro-cavity systems. \cite{mckeever2004deterministic,birnbaum2005photon, hijlkema2007single, wilk2007single, dayan2008photon} In most cases, the prototype system that has been widely explored is cavity quantum electrodynamics. The addition of an atom or quantum dot to a cavity makes it a highly nonlinear system and induces break down of the harmonicity of the energy levels of the cavity. This leads to the phenomenon of photon blockade, where only a single photon can occupy the cavity-atom or cavity-quantum dot system at a time.\cite{faraon2008coherent,birnbaum2005photon} Presence of one photon prohibits the simultaneous transmission of multiple photons through the cavity and a stream of one-by-one single photons at the output of the device from a coherent light input could be achieved. Therefore, photon blockade is characterized by observation of sub-Poissonian statistics of the output field. Photon blockade via creation of destructive quantum interference between transition pathways is also proposed by several authors.\cite{majumdar2012loss, liew2010single}

Recently, engineered quantum devices based on mechanical oscillators such as beam, cantilever etc. on micro and nano scale, have received tremendous attention.\cite{bowen2015quantum} Typical vibration frequencies of these mechanical oscillators lie in the range of kHz-GHz. Though controlling these systems electrically have been widely explored, it is only very recently, scientists have started investigating the coupling of these systems to electromagnetic field in the optical domain.\cite{aspelmeyer2014cavity} And this has opened up the emerging field of research called cavity optomechanics.\cite{bowen2015quantum} A typical optomechanical system consists of an optical cavity in which one of the end-mirrors is movable by the force exerted by light. Laser light entering the cavity exerts radiation pressure force, which deflects this mirror. Due to the mirror displacement, the length of the cavity is changed. Since the cavity resonance frequency depends on the length of the cavity, the cavity frequency is modified, thereby altering the intensity of the cavity field. This type of systems is intrinsically nonlinear due to the coupling between optical field and mechanical motion that can be realized using a driven cavity mode whose resonance frequency depends on the displacement of the mechanical degree of freedom. So far variety of optomechanical setups has been realized in practice, such as laser driven optical as well as microwave cavities, massive suspended mirrors, as well as micromechanical membranes, microdisks and nanobeams.\cite{aspelmeyer2014cavity} The recent surge of interest in the research of these systems can be attributed to the possibility of applications towards ground state cooling of mesoscopic oscillators, \cite{marquardt2007quantum, teufel2011sideband, sarma2016ground} entanglement of optical and mechanical modes, \cite{vitali2007optomechanical} nonclassical state generation, \cite{bose1997preparation, paternostro2011engineering} quantum state transfer between different modes \cite{tian2012adiabatic, dong2012optomechanical} etc. In particular, realization of photon blockade in optomechanical systems by virtue of the photon-phonon interaction induced nonlinearity has also been proposed.\cite{rabl2011photon,nunnenkamp2011single} However, one major challenge is that the effect of one photon tends to be very tiny. Towards this end, use of optomechanical photonic crystals could circumvent the weak nonlinearity due to a single photon. Optomechanical photonic crystals offer the privilege of added material-induced nonlinearity such as third-order nonlinear susceptibility, in addition to the inherent Kerr-type nonlinearity. It is worthwhile to mention that in cavity QED system, use of both second and third order nonlinearity to obtain single photon blockade is proposed by D. Gerace’s group.\cite{ferretti2012single, majumdar2013single} Photonic crystals offer the advantage of availability of diffraction-limited mode volumes and ultrahigh quality factors.\cite{notomi2010manipulating} These traits have been highly exploited towards obtaining effective optomechanical interaction in several works such as periodically patterned `zipper' system \cite{chan2009optical,eichenfield2009picogram}, single nanobeam structure confining both optical and mechanical vibration modes \cite{eichenfield2009optomechanical}, parallel photonic crystal
slabs \cite{notomi2006optomechanical}, double-beam nanomechanical resonator embedded in a photonic crystal slab cavity capable of highly efficient transduction of femtogram nanobeam resonators \cite{sun2012femtogram} etc. It may be noted that besides the generic optomechanical cavity system, few other configurations aiming at single-photon manipulation have been proposed such as: a quadratically coupled optomechanical cavity, \cite{liao2013photon,liao2014single,xie2016single} quantum-criticality-induced Kerr nonlinearities, \cite{lu2013quantum} application of pulsed laser drive, \cite{qiu2013single} a two-level system connected to the mechanical mode, \cite{wang2015tunable} optomechanical cavity coupled to two empty cavities, \cite{zhang2015controlling} coherent-feedback-induced effects \cite{liu2015coherent} etc. On the other hand, the prospective long distance communication of single-photon requires the workability around 1550 nm, i.e. the so-called telecommunication wavelength, as the photons generated should also have a wavelength that would give low loss and attenuation while traveling through an optical fiber. This issue has not been explored much in optomechanical systems. 

In this work, we demonstrate the possibility of realization of photon blockade at the telecommunication wavelength in an optomechanical photonic crystal cavity fabricated on a centrosymmetric medium, and using state-of-the-art nonlinearity parameters we show that photon blockade can be generated at weak driving condition. It should be noted that, unlike in bulk media  where the nonlinear susceptibility is very small at low input intensity, here it is significant. This is because, thanks to the recent advances in nanofabrication techniques, the weak nonlinearity is effectively enhanced to work even at the low-intensity level for diffraction-limited mode volumes. In this work, we analyze the antibunching properties of the photon field in terms of the photon statistics by applying a coherent weak laser drive. We find out the second-order photon correlation both numerically and analytically, and also check the robustness against decoherence effects like pure dephasing. We show that the proposed setup can be conveniently used as a single-photon blockade device for state-of-the-art nonlinear semiconductor material parameters. 
\section{\label{sec:level1}THEORETICAL FRAMEWORK}
Theoretically, the system can be modeled schematically as shown in Fig.1. We study an optomechanical cavity consisting of a centrosymmetric \cite{boyd2008nonlinear} medium, where there is a radiation-pressure induced coupling between a single optical mode and mechanical mode. For a dielectric material the nonlinear optical response to the applied electromagnetic field is given by:

\begin{align}\label{eq1}
\nonumber
D_i(\textbf{r},t) = \varepsilon_0 \varepsilon_{ij}(\textbf{\textbf{r}})E_j(\textbf{r},t) + \varepsilon_0[\chi_{ijk}^{(2)}(r)E_j(\textbf{r},t)E_k(\textbf{r},t)\\
+\chi_{ijkl}^{(3)}(\textbf{r})E_j(\textbf{r},t)E_k(\textbf{r},t)E_l(\textbf{r},t) +\dots].
\end{align}
Here, we are considering a centrosymmetric medium for which $\chi_{ijk}^{(2)}(\textbf{r})=0$. Considering the medium to be isotropic i.e. $\varepsilon_{ij}(\textbf{r})=\varepsilon(\textbf{r})$, the quantized electric and magnetic field operators for a single mode of the cavity field can be expressed as:
\begin{align}\label{eq1}
\textbf{E}(\textbf{r},t) = i\left(\frac{\hbar\omega_a}{2\varepsilon_0}\right)^{1/2}\left[a\frac{\alpha(\textbf{r})}{\sqrt{\varepsilon(\textbf{r})}}e^{-i\omega_at}-a^\dagger\frac{\alpha^*(\textbf{r})}{\sqrt{\varepsilon(\textbf{r})}}e^{i\omega_at}\right]
\end{align}
and 
\begin{align}
	\nonumber
\textbf{B}(\textbf{r}) = (-i/\omega_a)\nabla \times \textbf{E}(\textbf{r}), 
\end{align}
where $a$ ($a^\dagger$) is the annihilation (creation) operator and $\omega_a$ is the resonance frequency of the cavity field. $\alpha(\textbf{r})$ is the three-dimensional cavity
field profile that is normalized according to the condition $\int{|\alpha(\textbf{r})|^2d\textbf{r}} = 1$.
The second-quantized Hamiltonian for the electromagnetic field can be derived by using the classical expression for the time-averaged total energy density, $H_a =\frac{1}{2}\colon\int{\left[\textbf{E}(\textbf{r}) \cdot \textbf{D}(\textbf{r}) + \textbf{H}(\textbf{r}) \cdot \textbf{B}({\textbf{r}})\right] d\textbf{r}}\colon$, where magnetic field strength is given by $\textbf{H} = \textbf{B}/\mu_0$ for a non-magnetic medium. \cite{drummond1980quantum,sipe2004effective,ferretti2012single}  Here $\colon \colon$ denotes normal ordering.
%%%%%%%%%%%%%% FIGURE 1 %%%%%%%%%%%
\begin {figure}[!]
\begin {center}
\includegraphics [width =\linewidth]{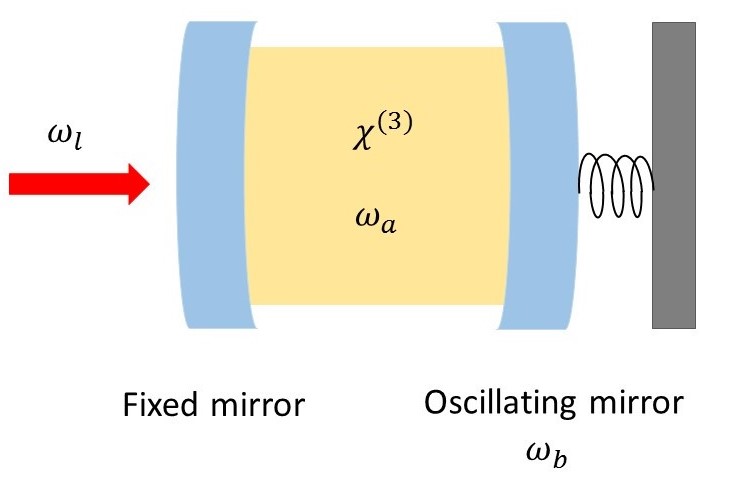}
\caption {(Color online) Schematic diagram of an optomechanical cavity containing a Kerr medium driven by a resonating coherent field.}
\label {fig1}
\end{center}
\end{figure}
%%%%%%%%%%%%%%%%%%%%%%%%%%%
%%%%%%%%%%%%%% FIGURE 2 %%%%%%%%%%%
\begin {figure}[!]
\begin {center}
\includegraphics [width =\linewidth]{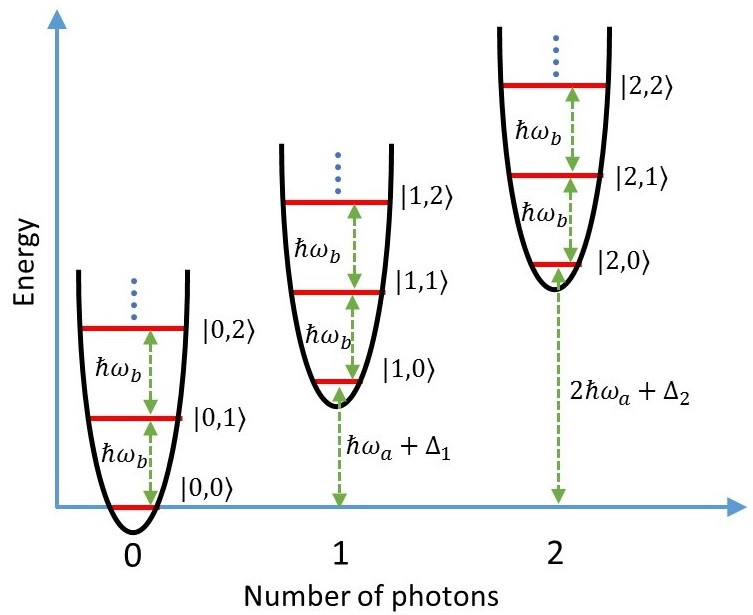}
\caption {(Color online) Energy level diagram of the system.}
\label {fig2}
\end{center}
\end{figure}
%%%%%%%%%%%%%%%%%%%%%%%%%%%
 Therefore, also considering the mechanical mode, the total Hamiltonian of the
system can be written as:
\begin{align} \label{eq2}
H_0=\hbar\omega_aa^\dagger a+\hbar\omega_bb^\dagger b+Ua^\dagger a^\dagger aa+\hbar ga^\dagger a\left(b+b^\dagger\right),
\end{align}
where, $\omega_b$ is the resonance frequency of the mechanical resonator and $b$ ($b^\dagger$) is the annihilation (creation) operator. $U$ is the nonlinear Kerr interaction strength that can be expressed in
terms of the classical electric field as:
\begin{align} \label{eq3}
	U=\frac{3(\hbar \omega_a)^2}{4\varepsilon_0}\int{\alpha^*_i(\textbf{r})  \frac{Re\{\chi^{(3)}_{ijkl}(\textbf{r})\}}{\varepsilon^2(\textbf{r})} \alpha^*_j(\textbf{r})\alpha_k(\textbf{r}) \alpha_l(\textbf{r})}d\textbf{r}.
\end{align}
Considering constant values of the average nonlinear susceptibilty, $Re\{\chi^{(3)}_{ijkl}(\textbf{r})\}= \bar{\chi}^{(3)}$ and relative dielectric permittivity, $\varepsilon(\textbf{r})= \bar{\varepsilon}_r$, the Kerr nonlinearity is given by:
\begin{align} \label{eq3}
	U=\frac{3(\hbar \omega_a)^2}{4\varepsilon_0}\frac{{\bar{\chi}}^{(3)}}{{\bar{\varepsilon}_r}^2}\int{|\alpha(r)|^4dr}= \frac{3(\hbar \omega_a)^2 D}{4\varepsilon_0 V_{eff}},
\end{align}
where $D=\frac{{\bar{\chi}}^{(3)}}{{\bar{\varepsilon}_r}^2}$ and $V_{eff}= \{\int{|\alpha(\textbf{r})|^4d\textbf{r}}\}^{-1}$ is the cavity mode volume. It is easy to see that by reducing the mode volume via suitable nanostructuring of the cavity system one could effectively enhance, unlike in ordinary bulk media, the strength of the nonlinear Kerr interaction even at the single-photon level.

The single-photon optomechanical coupling strength between the cavity
field and the mechanical resonator is denoted by $g$. Besides the Kerr nonlinearity, the interaction of the cavity field with the mechanical mode gives rise to another Kerr-type nonlinearity. To show it distinctly, we diagonalize the Hamiltonian by applying
Lang-Firsov unitary transformation given by $U=e^{-R}$ with $R = \frac{g}{\omega_b} a^\dagger a(b^\dagger - b)$. \cite{mahan2013many} We obtain the following transformed Hamiltonian: 
\begin{align} \label{eq3}
H^{\prime}_{0}=\left(\hbar\omega_a-U\right)a^\dagger a+\hbar\omega_bb^\dagger b+\left(U-\hbar\frac{g^2}{\omega_b}\right)n^2.
\end{align}
As can be seen from this Hamiltonian, the material nonlinearity as well as the optomechanical coupling leads to the photon energy level shift as shown schematically in Fig. 2. For zero to one photon transition, the energy difference is $\hbar\omega_a+\Delta_1= \hbar\omega_a-\hbar g^2/\omega_b$, whereas for one to two photon transition, the difference is $\hbar \omega_a + (\Delta_2-\Delta_1)= \hbar \omega_a+ ( 2U -3 \hbar g^2/\omega_b)$. Therefore, resonant transition to single-photon state `blocks' the absorption of a second photon because transitions to that level is detuned from resonance.

\section{\label{sec:level1} CALCULATION OF CORRELATION FUNCTION}
\subsection{\label{sec:level2} Numerical solution using Master equation approach}
The statistical properties of the cavity field can be described by the normalized zero-time-delay second-order correlation function given by
\begin{align} \label{eq4} 
	g^{(2)}(0)= \frac{\langle a^\dagger (t) a^\dagger (t) a(t) a(t)\rangle}{\langle a^\dagger (t) a(t)\rangle^2}.
\end{align}	
For photon blockade, we assume that the cavity is driven by a weak classical field with frequency
$\omega_l$. For convenience, we study the dynamics of such a system in a frame of reference rotating at the laser
frequency. The Hamiltonian of the driven system is given by:
\begin{align} \label{eq4}
	\nonumber	
	H_r=\hbar\Delta_aa^\dagger a+\hbar\omega_bb^\dagger b+Ua^\dagger a^\dagger aa\\
	+\hbar ga^\dagger a\left(b+b^\dagger\right)
	+\Omega(a^\dagger+a),
\end{align}
where $\Delta=\omega_a-\omega_l$ is the detuning of the cavity field and $\Omega$ is the coupling strength between the cavity field and the driving field.
The master
equation of the density operator $\rho$ for the driven system reads as:
\begin{align} \label{eq4}	
	\dot{\rho}=\frac{i}{\hbar}[\rho, H_r]+L_a(\rho)+L_b(\rho),
\end{align}
where $L_a(\rho)= \kappa n_a(a\rho a^\dagger+a^\dagger \rho a-a^\dagger a\rho-\rho a^\dagger a)+\frac{\kappa}{2}(2a \rho a^\dagger -a^\dagger a \rho -\rho a^\dagger a)$ and $L_b(\rho)= \gamma n_b(b\rho b^\dagger+b^\dagger \rho b-b^\dagger b\rho-\rho b^\dagger b)+\frac{\gamma}{2}(2b \rho b^\dagger -b^\dagger b \rho -\rho b^\dagger b)$ are the Liouvillian operators for the optical and mechanical modes respectively. $\kappa$ and $\gamma$ are the corresponding decay rates of the modes. $n_a$ and $n_b$ are the thermal photon and phonon numbers given by $n_i=\{exp[\hbar \omega_i/(k_BT)]-1\}^{-1}$. It is worth to be noted that due to the high frequency of optical radiation, thermal photon number is considered to be zero at low temperature.

%%%%%%%%%%%%%% FIGURE 3 %%%%%%%%%%%
\begin {figure}[!]
\begin {center}
\includegraphics [width =\linewidth, height = 0.6\linewidth]{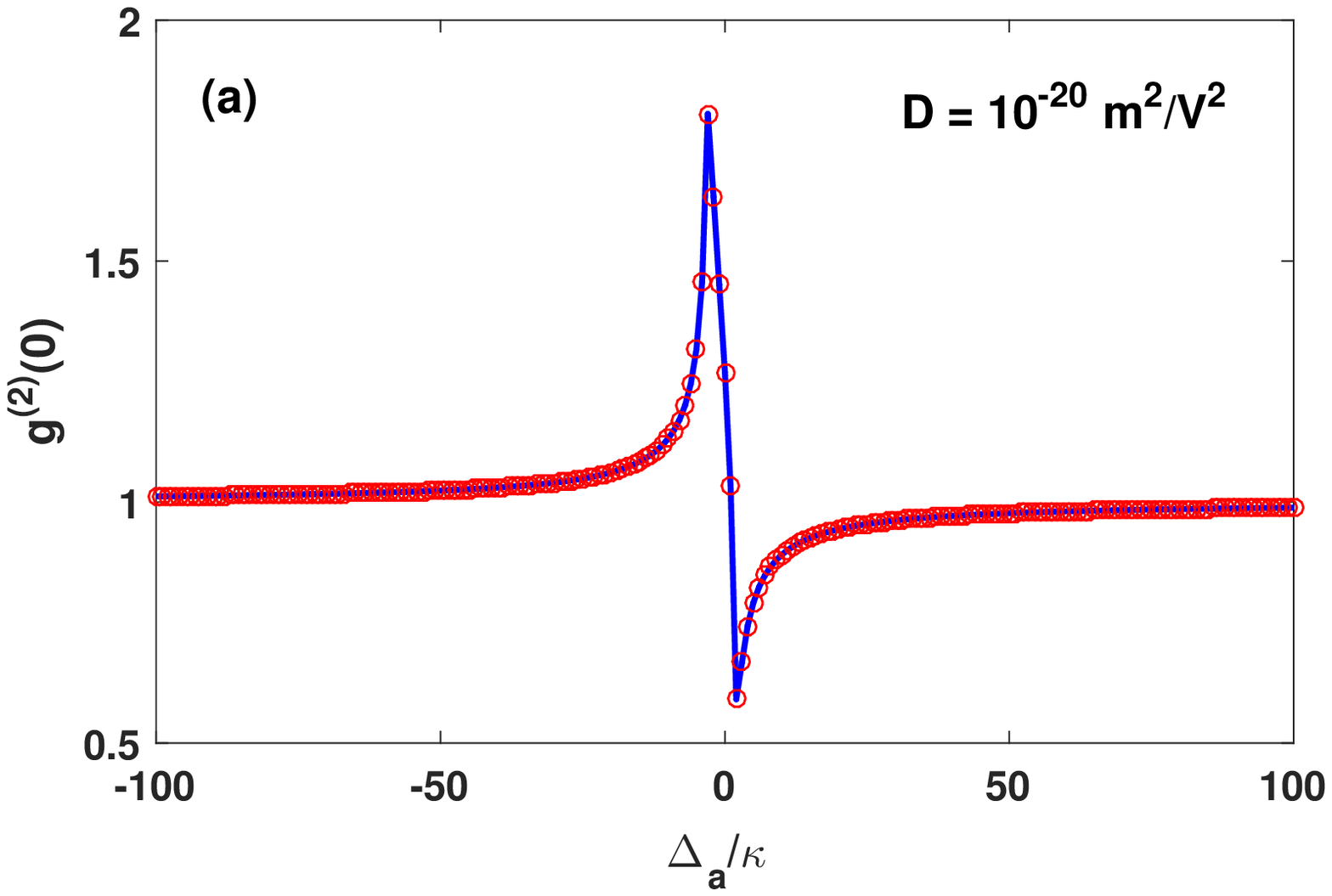}
\includegraphics [width =\linewidth, height = 0.6\linewidth]{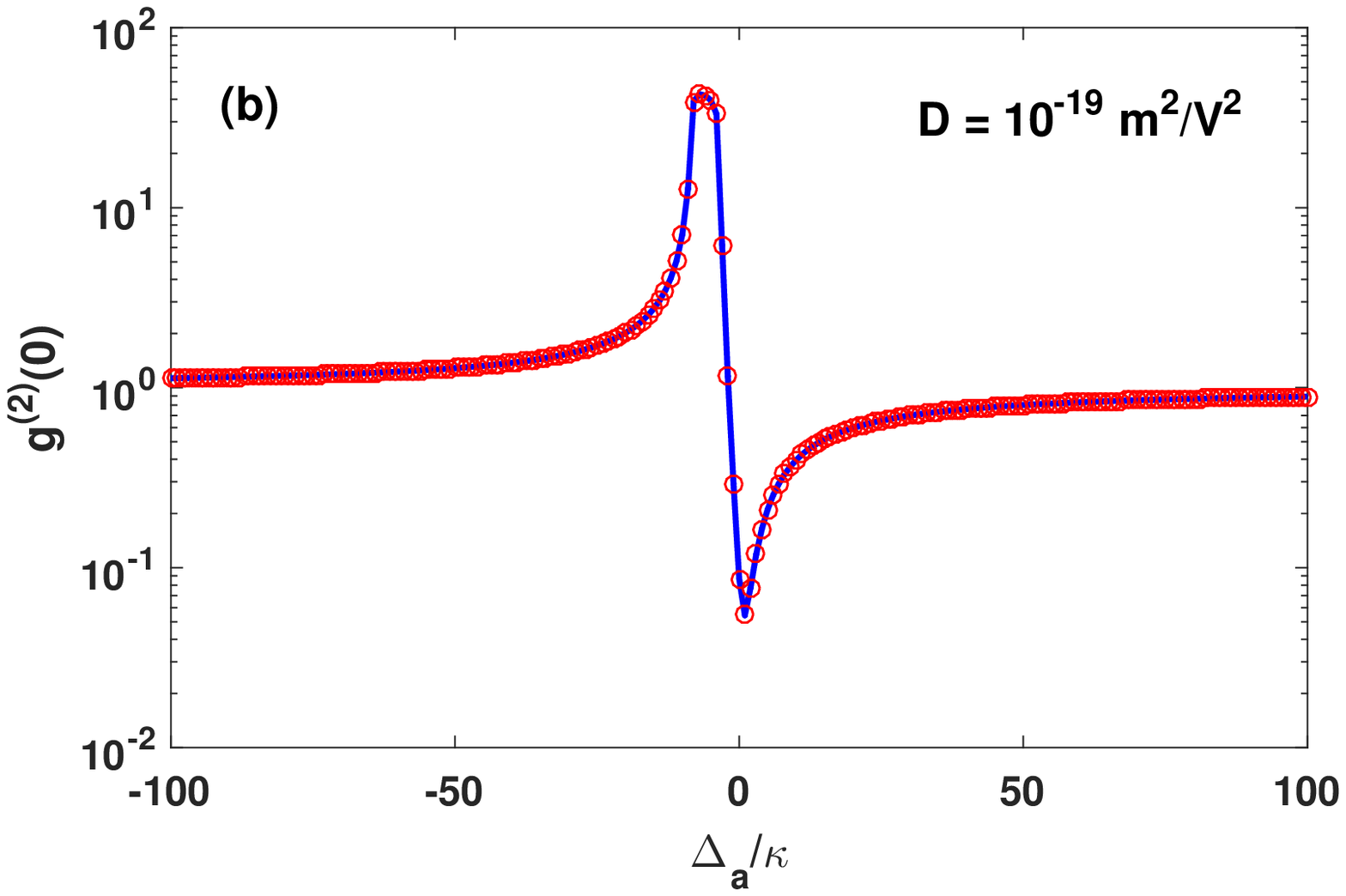}
\includegraphics [width =\linewidth, height = 0.6\linewidth]{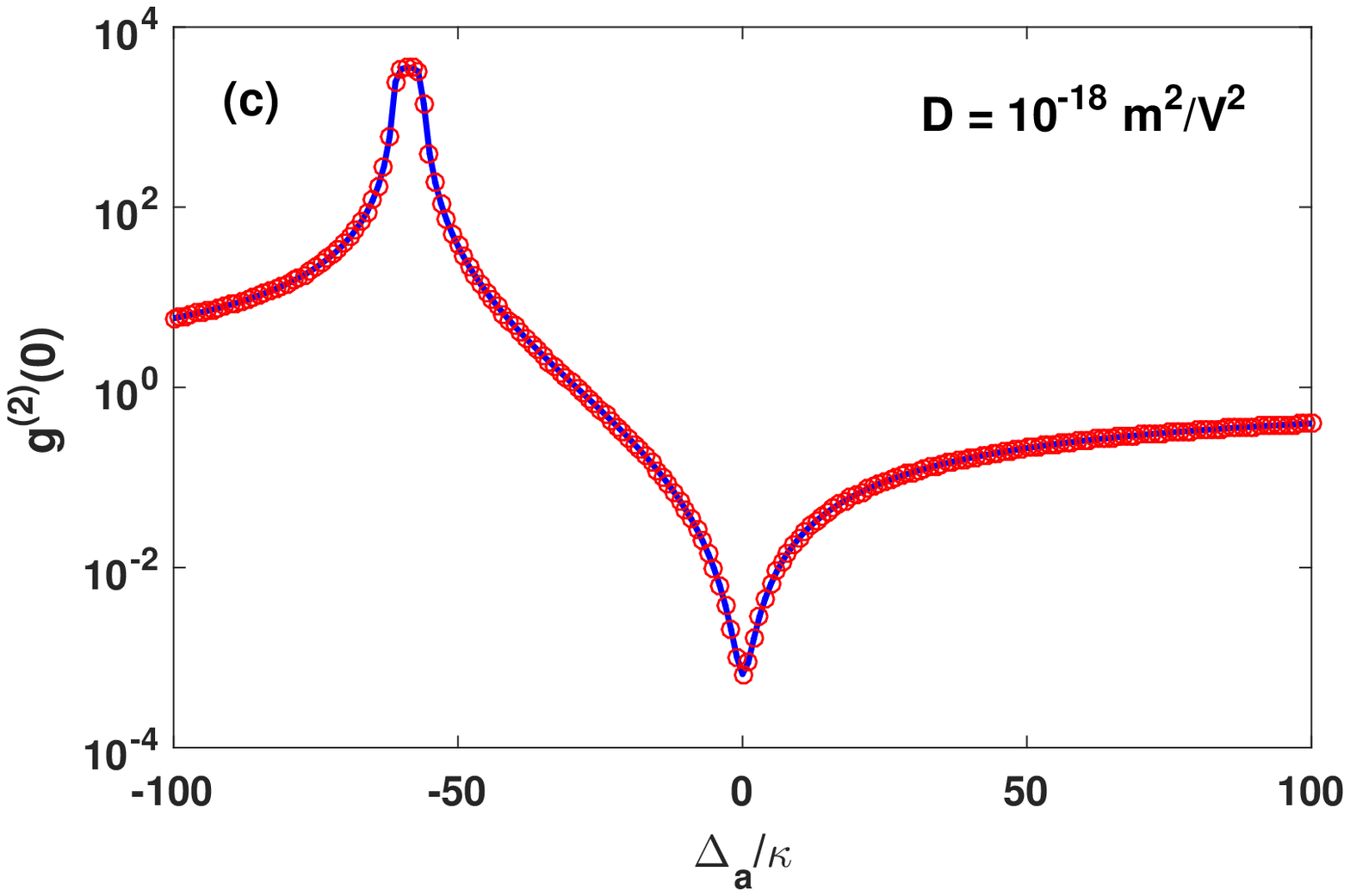}
\includegraphics [width =\linewidth, height = 0.6\linewidth]{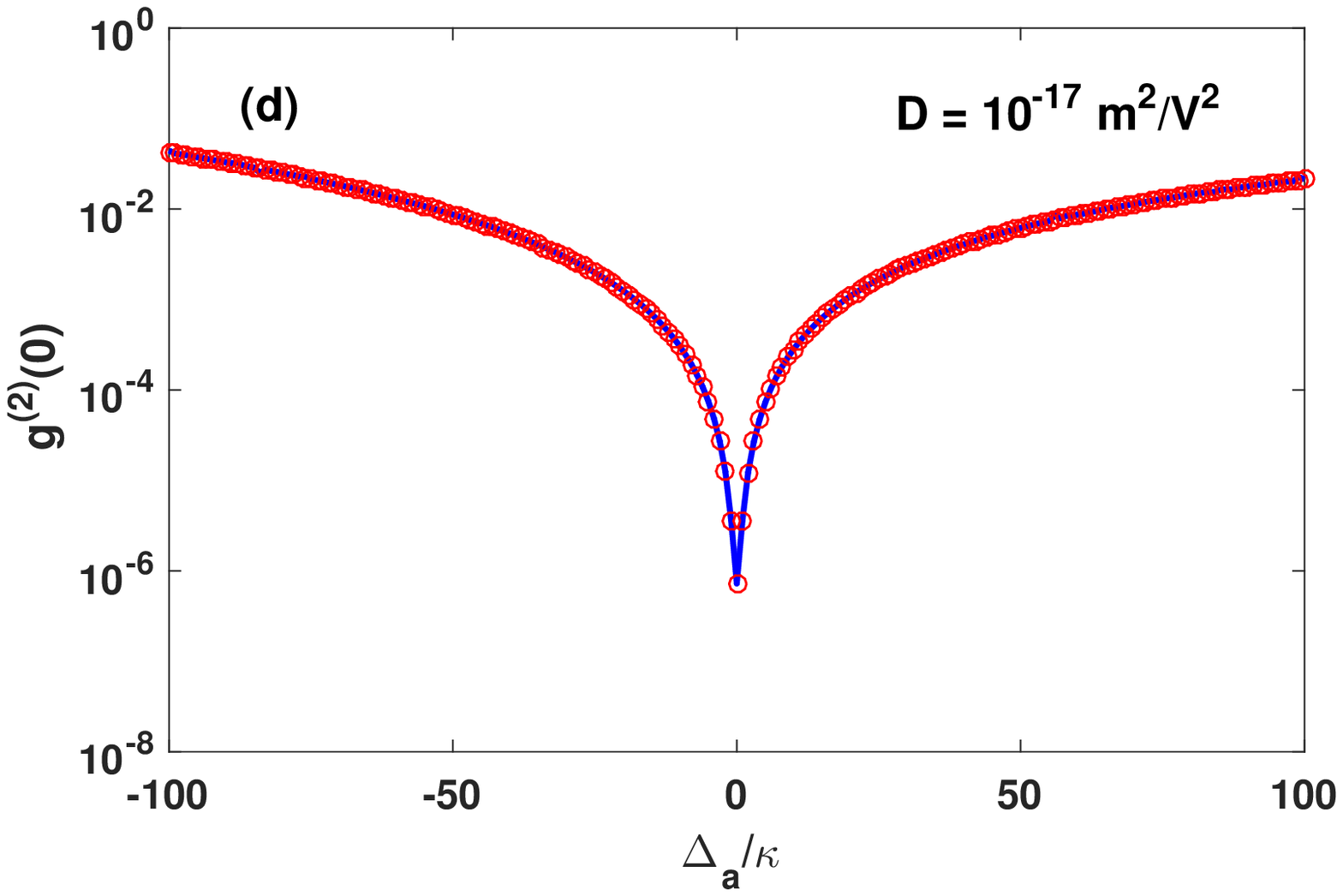}
\caption {(Color online) Numerical (solid line, blue) and analytical (symbols, red) solutions for zero-time-delay second-order correlation function for different values of the ratio $D = \chi^{(3)}/{\bar{\varepsilon}_r}^2$, as functions of normalized cavity detuning. $V_{eff}$ is taken to be 0.01 $\mu m^3$. Parameters are $\omega_a/2\pi=205.6 \,THz$, $\omega_b/2\pi= 9.5 \,GHz$, $Q=5.1\times10^7$, $\Omega=0.01\kappa$ with $\kappa= \omega_a/Q$; $T=0$, $\gamma=0.001\kappa$, $g/2\pi=292 \,kHz$.}
\label {fig3}
\end{center}
\end{figure}
%%%%%%%%%%%%%%%%%%%%%%%%%%%
The steady-state value of $g^{(2)}(0)$ can be found numerically by solving the master equation and from the steady state density matrix operator as:
\begin{align} \label{eq4}	
	g^{(2)}(0)= \frac{Tr(\rho a^\dagger a^\dagger a a)}{[Tr(\rho a^\dagger a)]^2}.
\end{align}

\subsection{\label{sec:level1} Analytical solution}
 In weak driving condition $(\Omega\ll\kappa)$ and for low temperature, the photon number and phonon number Hilbert space can be truncated to low values. Considering a basis state, $|m,n\rangle$ in Hilbert space, where  $m$ and $n$ are photon and phonon numbers respectively, we truncate the states according to $m+n\leq2$. Then the state of the system can be expressed as: \cite{bamba2011origin}
\begin{align} \label{eq4}
\nonumber	
	|\psi\rangle= C_{00}|0,0\rangle+C_{01}|0,1\rangle+C_{10}|1,0\rangle+C_{11}|1,1\rangle\\
+C_{20}|2,0\rangle+C_{02}|0,2\rangle,
\end{align}
where $C_{mn}$ are the amplitudes of the quantum states and the corresponding occupation probability is given by $|C_{mn}|^2$.
Using this, the second-order degree of coherence can then be given as:
\begin{align} \label{eq4}	
	g^{(2)}(0)= \frac{2|C_{20}|^2}{(|C_{10}|^2+|C_{11}|^2)^2}.
\end{align}
The coefficients $C_{mn}$ can be obtained by solving the Schr\"{o}dinger equation:
\begin{align} \label{eq4}	
	i\hbar\frac{d|\psi\rangle}{dt}=H_r'|\psi\rangle,
\end{align}
where $H_r'$ is the effective non-Hermitian Hamiltonian that takes into account the dissipations in the system:
\begin{align} \label{eq4}
\nonumber	
    H_r'= \hbar \Delta_a' a^\dagger a+ \hbar \omega_b' b^\dagger b+U a^\dagger a^\dagger a a\\
    +\hbar g a^\dagger a (b+b^\dagger)+\Omega(a^\dagger +a).	
\end{align}
with $\Delta_a'= \Delta_a- i \kappa/2$ and $\omega_b'= \omega_b- i\gamma/2$.
In the steady state, the coefficients are calculated by putting $\frac{d|\psi\rangle}{dt}=0$ and for very weak pump, considering $C_{00}\approx 1$. The coefficients to be used for the calculation of $g^{(2)}(0)$ are obtained as:
\begin{align} \label{eq4}
\nonumber	
	C_{20}= -\frac{\Omega}{\sqrt{2}(\Delta_a'+\frac{U}{\hbar})}C_{10}\\
    C_{11}= -\frac{g}{\Delta_a'+\omega_b'-\frac{\Omega^2}{\omega_b'}}C_{10}\\
\nonumber
    C_{10}= \frac{\Omega}{\frac{g^2}{\Delta_a'+\omega_b'-\frac{\Omega^2}{\omega_b'}}-(\Delta_a'-\frac{\Omega^2}{\Delta_a'+\frac{U}{\hbar}})}
\end{align}
It is worth to be noted that in absence of the optomechanical coupling, i.e. for the case, $g=0$, the zero-time delay second-order correlation function is correctly approximated to the value:
\begin{align} \label{eq4}	
	g^{(2)}(0)= \frac{1+4\frac{\Delta_a^2}{\kappa^2}}{1+4\frac{(\Delta_a+\frac{U}{\hbar})^2}{\kappa^2}}.
\end{align}
The results of Eq. (12) together with (15) will be compared with the numerical
results in the low-pumping and low temperature conditions in the following section.

%%%%%%%%%%%%%% FIGURE 4 %%%%%%%%%%%
\begin {figure}[!]
\begin {center}
\includegraphics [width =\linewidth]{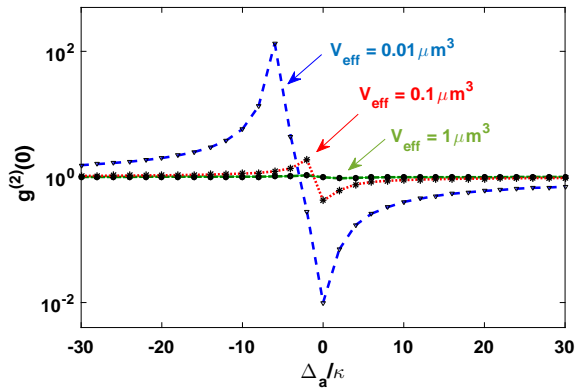}
\caption {(Color online) Zero-time-delay second-order correlation function for different values of effective confinement volume of the cavity field, $V_{eff}$. The nonlinearity ratio $D=\chi^{(3)}/{\bar{\varepsilon}_r}^2=10^{-19} m^2/V^2$. Other parameters are same as in Fig. 3. Numerical results are shown as lines, whereas analytical results are shown in symbols and both matches precisely. The green, red and blue lines correspond to $V_{eff} = 1 \mu m^3$, $0.1 \mu m^3$ and $0.01 \mu m^3$ respectively.}
\label {fig4}
\end{center}
\end{figure}
%%%%%%%%%%%%%%%%%%%%%%%%%%% 
%%%%%%%%%%%%%% FIGURE 5 %%%%%%%%%%%
\begin {figure}[!]
\begin {center}
\includegraphics [width =\linewidth]{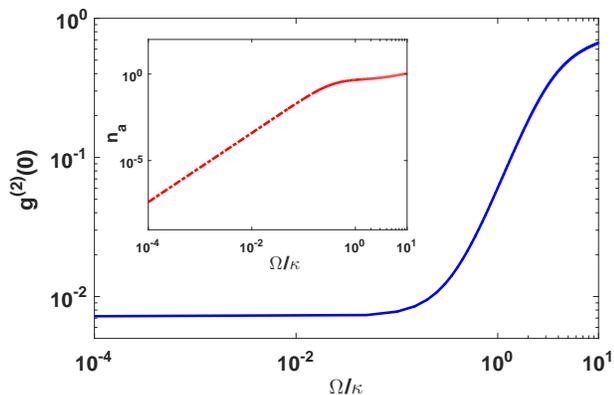}
\caption {(Color online) Variation of zero-time-delay second-order correlation function and photon number (inset) as functions of pumping rate. $\Delta_a = 0$, $D=10^{-19} m^2/V^2$, other parameters are same as in Fig. 4.}
\label {fig5}
\end{center}
\end{figure}
%%%%%%%%%%%%%%%%%%%%%%%%%%%

\section{\label{sec:level1} PHOTON BLOCKADE}
Photon blockade is one of the most striking demonstrations of the fact that, owing to photon-phonon interactions, an optomechanical cavity act effectively as a nonlinear medium. The steady-state value of the normalized zero-time-delay second-order correlation $g^{(2)}(0)$, discussed in the previous section, could be used as a figure of merit for quantifying single photon blockade behavior \cite{fox2006quantum}. The value of second-order correlation function of a system is a statistical indicator of the nonclassicality of the system. $g^{(2)}(0) > 1$ indicates super-Poissonian statistics which is a classical effect, whereas $g^{(2)}(0) < 1$ corresponds to sub-Poissonian statistics of the cavity field, which is a nonclassical effect. In this section we discuss this statistical property of the system and its dependence on the system parameters. The dips of $g^{(2)}(0)$ showing sub-Poissonian behavior characterize the photon blockade process. In this process due to the anharmonicity of the energy levels, only single-photon transition takes place. The peaks of $g^{(2)}(0)$ showing super-Poissonian behavior signifies multiphoton transition owing to photon tunneling which is analogous to electron-tunneling in solid-state systems. \cite{chiao1991analogies, xu2013photon}

In Fig. 3, $g^{(2)}(0)$ is shown as a function of normalized cavity detuning for different values of the ratio $D=\chi^{(3)}/{\bar{\varepsilon}_r}^2$. For calculations, we consider a state-of-the-art experimental photonic optomechanical cavity (Si) system with $\omega_a/2\pi=205.6 \,THz$ for $\lambda_a\simeq 1.5 \mu m (=1.459 \mu m)$ in the typical near-infrared range; $\omega_b/2\pi= 9.5 \,GHz$, $\kappa= \omega_a/Q_a$ ($Q_a=5.1\times 10^7$), $\gamma=0.001 \kappa$, $g/2\pi=292 \,kHz\,$. \cite{safavi2010design} Here, we have considered the parameters in weak single-photon optomechanical coupling regime, i.e. $g<\kappa$. For weak pumping and low temperature limit, we consider $\Omega=0.01 \kappa$ with $T=0 K$. The values of $\chi^{(3)}$ and $\bar{\varepsilon}_r$ are material-dependent. From the literature, we find the $\chi^{(3)}$ tensor elements for typical semiconductors as of the order of $\chi^{(3)} \approx 10^{-18} - 10^{-19} m^2/V^2\,$, and for some nanoparticle-doped glasses, of the order of $10^{-16} m^2/V^2\,$ in near-infrared. \cite{dinu2003third, bristow2007two, hon2011third, razzari2006excited, martinez2010ultrafast,ganeev2006nonlinear} The $\bar{n}_r$ ($\bar{n}_r=\sqrt{\bar{\varepsilon}_r}$) values vary typically from 2 to 4. \cite{boyd2008nonlinear} Considering diffraction-limited confinement volumes, i.e., $V_{eff} \sim (\lambda_a/2\bar{n}_r)^3$ for Si based system, a realistic value $V_{eff} = 0.01 \,\mu m^3$ is assumed. The plots show strong antibunching near zero cavity detuning. It is worth to be noted that even for lower values of nonlinear susceptibility among the available nonlinear materials: $D=10^{-20} m^2/V^2$, there is strong antibunching effect. In Fig. 4, the influence of confinement volume $V_{eff}$ on $g^{(2)}(0)$ is demonstrated. For order of magnitude value of $V_{eff}= 0.01 \,\mu m^3$, there is strong antibunching near zero cavity detuning. With higher values of $V_{eff}$, the value of $g^{(2)}(0)$ merges towards Poissonian statistics. In Fig.s 3 and 4, we show both numerical and analytical solutions and both match perfectly.  There is also photon tunneling effect at some negative detunings. This occurs due to the fulfillment of two-photon resonance condition by the incoming laser, since at negative detuning the laser frequency is higher than the single-photon resonance frequency.

%%%%%%%%%%%%%% FIGURE 6 %%%%%%%%%%%
\begin {figure}[!]
\begin {center}
\includegraphics [width =\linewidth]{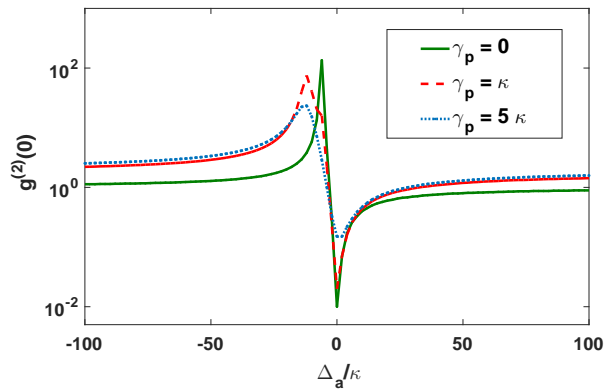}
\caption {(Color online) Variation of zero-time-delay second-order correlation function with respect to pure dephasing rates of the photon field.}
\label {fig6}
\end{center}
\end{figure}
%%%%%%%%%%%%%%%%%%%%%%%%%%%

The choice of working in the weak pumping strength regime is justified in Fig. 5. We show the variation of $g^{(2)}(0)$ and also the variation of average number of cavity photons $n_a$ (inset) as function of pump amplitude. On increasing the pump amplitude, $g^{(2)}(0)$ approaches towards Poissonian characteristics and there is also increase in the average photon number. 

We now briefly analyse the effect of pure dephasing, which is a detrimental environmental effect perturbing the photon statistics \cite{majumdar2012probing, englund2012ultrafast, auffeves2010controlling}, polarization \cite{majumdar2011phonon}, linewidth \cite{majumdar2010linewidth, yamaguchi2008photon}, and transmittance. This can arise due to thermal instability, coupling to other modes and also due to instability of the laser pump. In order to inspect the effect of pure dephasing on photon blockade, we add another Lindblad term  of the form $L_p(\rho) = \frac{\gamma_p}{2}(2a^\dagger a \rho a^\dagger a - (a^\dagger a)^2 \rho - \rho (a^\dagger a)^2)$, in the master equation. $\gamma_p$ is the pure dephasing rate for the cavity mode. Fig. 6 shows the second-order correlation function
$g^{(2)}(0)$ with different pure dephasing rates. As can be seen from the figure, the qualitative nature of the photon
blockade is unchanged with $\gamma_p$, still showing antibunching at near-zero cavity detuning. For order of magnitude value of pure
dephasing rate $\gamma_p= \kappa$, the
antibunching dip is affected marginally. With increase in pure
dephasing rate, the antibunching dip is lessened. However, for a typical pure dephasing rate of the order of $0.1 \kappa$, the device can safely be treated as a robust single photon source with photon blockade.
 
\section{\label{sec:level1}CONCLUSION}	
In summary, we propose the use of optomechanical photonic crystals, comprising of centrosymmetric material, as an effective arrangement for generating single-photon states via photon blockade. Our proposal is motivated by state-of-the-art experimental advancements in nanofabrication techniques and the low mode-volume offered by photonic crystal structures.  We investigate the photon blockade effect at weak resonant driving by calculating the zero-time-delay second order correlation function, $g^{(2)}(0)$, of the cavity photon, by solving the Schr\"{o}dinger equation analytically. This is again confirmed by the numerical results obtained from solving the quantum master equation. Considering realistic parameters, we have shown that strong photon antibunching could be achieved at weak single-photon optomechanical coupling regime and even for lower values of the nonlinear susceptibility parameter. The setup could conveniently be used as a single-photon generator at telecommunication wavelength, which is necessary for long-distance quantum communication. Also, the antibunching effect is found to be robust for typical range of pure-dephasing induced decoherences. Our proposal might pave the way for making integrated optomechanical single-photon sources at telecommunication wavelength.

\begin{flushleft}
\textbf{Acknowledgements}
\end{flushleft}
B. Sarma would like to thank MHRD, Government of India for a research fellowship.

\end{document}